\documentclass[%
 reprint,
amsmath,amssymb,
aps, 
]{revtex4-2}

\usepackage{graphicx}
\usepackage{dcolumn}
\usepackage{bm}
\usepackage{comment}

\begin{document}

\preprint{PRN/123-QED}
\title{
Photoluminescence efficiency droop in Perovskites
}

\author{Pradeep R. Nair}
 \email{prnair@ee.iitb.ac.in}
\author{Karthik Raitani}
\affiliation{ Department of Electrical Engineering,
 Indian Institute of Technology Bombay, Mumbai, India\\
}%




\date{\today}

\begin{abstract}
The commercialization prospects of perovskite light emitting diodes depend on its luminescence efficiency under large carrier densities. The decrease in luminescence efficiency under such high injection conditions could lead to an undesired increase in power consumption with associated degradation and stability concerns. Here, through detailed modeling of thermal transport and carrier generation-recombination, we unravel the physical mechanisms that cause luminescence droop under high injection conditions. We show that self-heating leads to a reduction in the radiative recombination (both bimolecular and excitonic). The resultant increase in non-radiative recombination and hence the thermal dissipation acts as a positive feedback mechanism that leads to efficiency droop in perovskites. Our model predictions, well supported by experimental results, could be of broad interest towards the degradation-aware thermal design of perovskite optoelectronics and stability.
\end{abstract}

\maketitle


\section*{Introduction}
Perovskite based light emitting diodes (LEDs) have achieved excellent conversion efficiencies \cite{Yoo2021,Shen2024,Sun2023,Kim2022,longi,Li2024}. The reported record external quantum efficiency (EQE) for 3D perovskite LEDs is about 32\% (with a corresponding Photoluminescence Quantum Efficiency, PLQE, of 96\%) \cite{Li2024}. However, EQE is not the only parameter of interest for LEDs. It is also equally important to achieve significant radiance under low power consumption. Hence, it is essential that such excellent EQE is demonstrated for large carrier densities. Equivalently, it is desirable that near ideal PLQE is demonstrated under large illumination intensities.  However, experimental reports indicate that PLQE decreases at high injection levels \cite{Li2024}. This reduction in the efficiency of radiative recombination (known as 'droop' in III-V semiconductors \cite{GaNdroop,droop_challenges}) is detrimental to practical applications, as it increases the power consumption required to achieve the desired radiance.\\

\begin {figure} [h]
  \centering
    \includegraphics[width=0.25\textwidth]{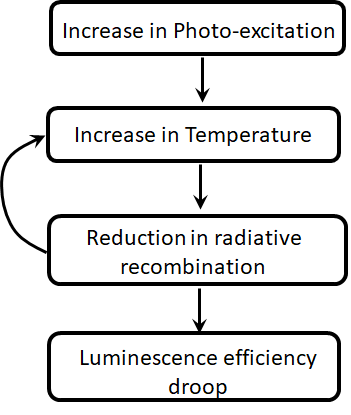}
    \caption{\textit{ Schematic illustrates the positive feedback mechanism involved in the luminescence efficiency droop in Perovskites.     
     }}
\label{schematic}
\end{figure}

Recent reports indicate that excitonic effects are significant in the improved performance of perovskite LEDs \cite{Simbula,Zhao2023,Mariano2020}. However, its effectiveness in high level injection conditions is yet to be well established. For practical applications, it is important that the factors which contribute to efficiency droop are well delineated and the device design is optimized to reduce the same. In this regard, here we show that the luminescence efficiency droop in perovskites is due to an interesting positive feedback mechanism between an increase in sample temperature and an associated  reduction in radiative recombination (see Fig. \ref{schematic}). The same is unravelled through a self consistent model which accounts for heat flux balance along with carrier   generation-recombination. Through this model, we quantify the influence of self-heating towards the reduction in luminescence efficiency and identify the demands on device design to achieve high radiance under low power consumption. Below, we describe the analytical model.

\section*{Temperature dependent PLQE}
The reduction in the relative contribution of radiative recombination and hence the PLQE is often attributed to increased Auger recombination. Under steady state photo-illumination, we have
\begin{equation}
     G = k_1n+k_2n^2+k_3n^3 \\
    \label{eq:sim_rate}
\end{equation}
where $n$ is the free carrier density and $G$ is the carrier generation rate due to incident photons. The parameters $k_1$, $k_2$, and $k_3$ denote the monomolecular, bimolecular, and Auger recombination processes \cite{stranks_PRA,Hossain2024}. Excitonic effects also contribute to the radiative recombination  \cite{stranks_PRA,nair2024_exciton}. The carrier generation rate is related to the intensity ($I$) of incident illumination as $I=GWhc/\lambda$, where $h$ is the Planck's constant, $c$ is the velocity of light, $\lambda$ is the wavelength of incident illumination, and $W$ is the sample thickness. Accordingly, the PLQE is given as
\begin{equation}
     PLQE = \frac{k_2n^2}{G}=\frac{k_2n^2}{k_1n+k_2n^2+k_3n^3}  \\
    \label{eq:sim_PLQE}
\end{equation}

The variation of PLQE with $G$ is as follows: For small $G$ and hence small $n$, monomolecular process is the dominant recombination mechanism. Accordingly, we have $G \approx k_1n$ andc hence we expect the PLQE to vary linearly with $G$. For large $G$, radiative process might be the dominant recombination mechanism which leads to $G\approx k_2n^2$. Under such conditions, eq. \ref{eq:sim_PLQE} predicts that the PLQE is rather independent on $n$. For much larger $G$, Auger recombination becomes significant with $G \approx k_3n^3$. Under such regime, eq. \ref{eq:sim_PLQE} indicates that PLQE varies as $G^{-1/3}$. Rather the PLQE decreases with $G$  under high illumination intensities. \\

The PLQE decreases from its peak value once the Auger recombination becomes significant, as compared to the radiative recombination. For a given set of parameters $k_1$, $k_2$, and $k_3$, eq. \ref{eq:sim_PLQE} indicates that the normalized PLQE decreases from 90\% to 50\% as the $n$ increases by a order of magnitude (i.e., from $0.1k_2/k_3$ to $k_2/k_3$). Over such large carrier densities, the radiative and Auger processes dominate the recombination. Hence an order of magnitude change in $n$, typically requires more than 2 orders of magnitude change in $G$. However, experimental results indicate that PLQE decreases much more rapidly with $G$ for large illumination intensities while the maximum PLQE is close to ideal limits \cite{Li2024}. Hence, it is evident that other physical mechanisms, in addition to the familiar increase in Auger recombination, could influence the PLQE at large illumination intensities. \\

\begin {figure} [h]
  \centering
    \includegraphics[width=0.45\textwidth]{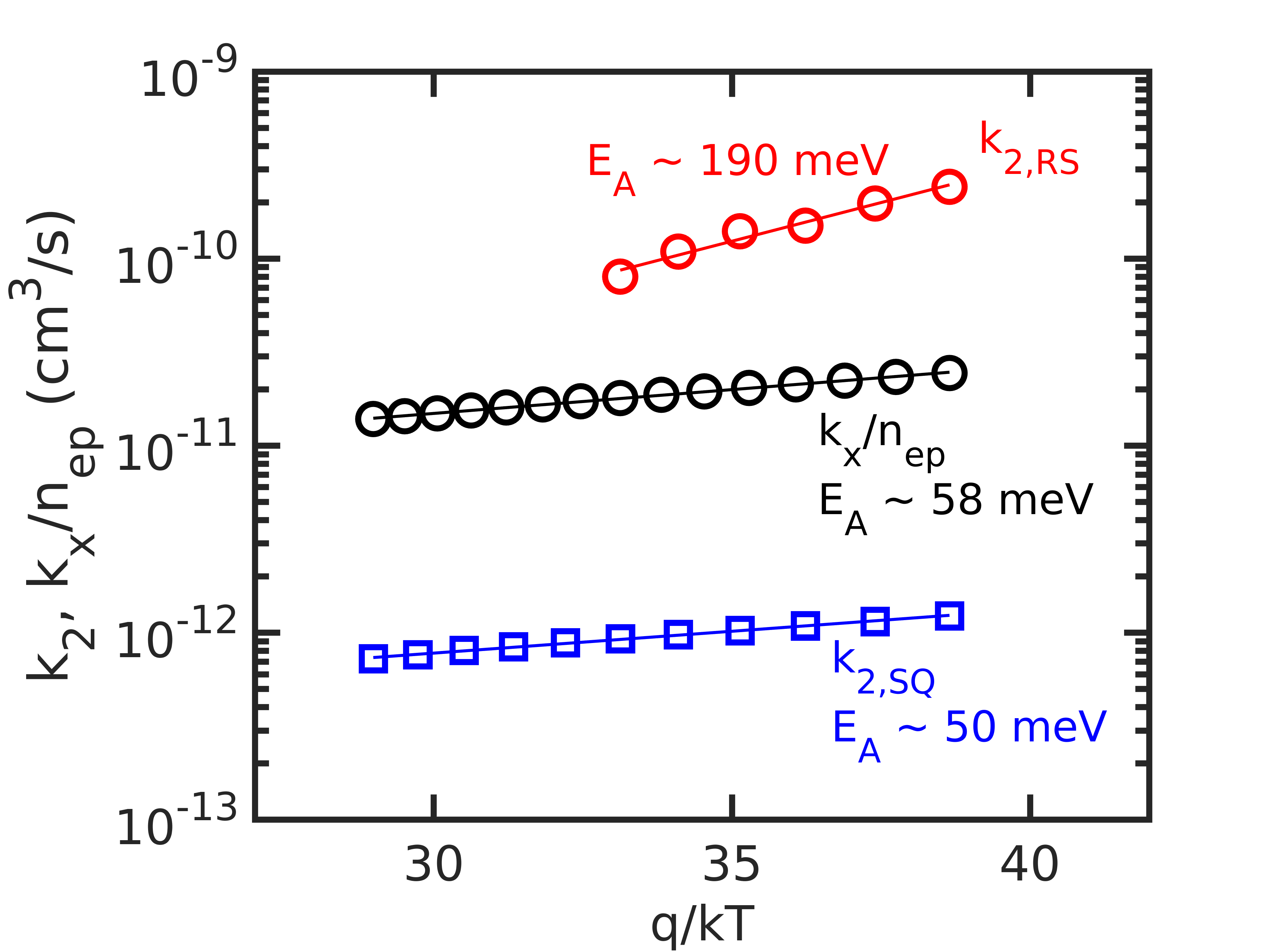}
    \caption{\textit{Variation of radiative recombination rate with temperature. The data set 'SQ' correspond to detailed balance analysis with unity absorption coefficient (i.e., eq. \ref{eq:k2_rad}). The results marked as 'RS' corresponds to estimates using van Roosbroeck-Shockley model (i.e., eq. \ref{eq:k2_SR}). Black symbols denote excitonic contribution to radiative recombination ($k_x/n_{ep}$, see eq. \ref{eq:ex}). The lines indicate fit to the equation \ref{eq:k2_T} and the back extracted $E_A$ for each case is as indicated. 
     }}
\label{k2_T}
\end{figure}

An important but yet to be appreciated (i.e., in the perovskite optoelectronics community) contributing factor towards luminescence efficiency droop is the influence of temperature. The sample under illumination could be at elevated temperatures and each of the parameters ($k_1$, $k_2$, and $k_3$) are temperature dependent. The parameter $k_2$ in eqs. \ref{eq:sim_rate}-\ref{eq:sim_PLQE} represents the effective bimolecular recombination rate. This has contribution from excitonic components \cite{stranks_PRA,nair2024_exciton} and is given as
\begin{subequations}
\begin{align}
         k_2 &= k_{2,rad}+\frac{k_x}{n_{ep}}\\
      n_{ep} &=(2\pi\mu kT/h^2)^{3/2}e^{-E_b/kT}   
\end{align}
    \label{eq:ex}
\end{subequations}
where $\mu$ is the reduced mass, $kT$ is the thermal energy, and $h$ is the Planck's constant. The detailed balance theory and Shockley-Queisser (SQ) analysis  \cite{SQ} indicate that the radiative recombination rate is influenced by the black body spectrum ($\Phi_{BB}$). Accordingly, for a sample of thickness $W$ and intrinsic carrier concentration $n_i$, an estimate for the radiative recombination rate \cite{SQ,bolink_eqeel} is 
\begin{equation}
     k_{2,SQ} = \frac{\int_{E_{g}}^{\infty} \Phi_{BB}(E)dE}{n_i^2W}  \\
    \label{eq:k2_rad}
\end{equation}
where $E$ is the energy. Given $\int_{E_{g}}^{\infty} \Phi_{BB}(E)dE \propto e^{-E_g/kT}$ and $n_i^2 \propto e^{-E_g/kT}$, we expect a weak temperature dependence for $k_{2,SQ}$ - an inference supported by the numerical estimates shown in Figure \ref{k2_T}. However, eq. \ref{eq:k2_rad} assumes all energetic photons are absorbed in the sample. In reality, the absorption coefficients depend on $E$ and are temperature dependent \cite{Raja_optical}. Accordingly, an estimate for the radiative recombination rate using the van Roosbroeck-Shockley (RS) model \cite{Roosbroeck} is
\begin{equation}
     k_{2,RS} = \frac{\int \alpha (E) \Phi_{BB}(E)dE}{n_i^2}  \\
    \label{eq:k2_SR}
\end{equation}
where $\alpha(E)$ is the probability of absorption of a photon with energy $E$. Using the recent reports on the temperature dependent refractive indices \cite{Raja_optical}, numerical estimates for $k_{2,SR}$ using eq. \ref{eq:k2_SR} is plotted in Fig. \ref{k2_T}. Due to $\alpha$, we find that $k_{2,RS}$ has a much stronger temperature dependence than $k_{2,SQ}$ (see Fig. \ref{k2_T}).\\

Equation \ref{eq:ex}b describes the temperature dependence of the excitonic component ($k_x/n_{ep}$).  As $T$ increases, $n_{ep}$ increases which contributes to a decrease in $k_2$. Similarly, $E_b$ directly influences the temperature depedence of $k_2$. In addition to $n_{ep}$, $k_x$ could also be temperature dependent (in accordance with von Roosbroeck-Shockley model \cite{Roosbroeck,Davies2018}). The temperature dependence of the excitonic component, assuming $k_x$ to be temperature independent, is plotted in Fig. \ref{k2_T}.\\

The results in Fig. \ref{k2_T} indicate that the temperature dependent radiative recombination rate (SQ, RS, or exciton) can be conveniently represented as
\begin{equation}
    k_2(T)\propto k_{2,F}e^{E_A/kT} \\
    \label{eq:k2_T}
\end{equation}
over a temperature range of interest ($300-350 \,\textnormal{K}$). The corresponding $E_A$ values are as indicated in Fig. \ref{k2_T}. As discussed, $E_A$ for the SQ model is comparable with the excitonic contribution (with $E_b =0.014 \mathrm{eV}$, see Suppl. Methods for the parameters used). However, the absorption coefficient seems to be the dominant factor that contributes to the temperature dependence of $k_2$. Indeed, experimental reports  on the transient absorption spectroscopy of perovskites \cite{Davies2018} also report similar trends for $k_2$ - in broad agreement with our calculations.\\ 

\section*{Self heating and Positive feedback}

Under illumination, a part of the incident energy is always dissipated in the sample.  For each incident photon which is absorbed, the generated free carriers relax to the bottom of the bands before they recombine through various processes. This energy loss is well known as the thermalization loss \cite{hirst_loss}. Further, monomolecular as well as Auger recombination are non-radiative in nature and leads to thermal dissipation (known as recombination loss). In fact, except for the radiative recombination and hence the photoluminescence, the entire incident power is dissipated in the sample which leads to an increase in its temperature. \\

Detailed balance for the heat flux indicates that the net heat flow should be zero under steady state conditions. Mathematically, using a lumped component model, this can be represented as
\begin{subequations}
\begin{align}
     G_{T}(T-T_{amb}) & = I - k_2n^2WE_g \\
          T & =T_{amb}+ \frac{I}{G_T}(1 - \frac{E_g}{E_{ph}} PLQE)
          \end{align}
    \label{eq:thermal}
\end{subequations}
where $G_{T}$ denotes the effective thermal conductance, $T$ is the sample temperature, and $T_{amb}$ is the ambient temperature. The LHS  of eq. \ref{eq:thermal}a denotes the heat outflux from the sample (per unit area), while the first term on the RHS denotes the incident power intensity (i.e., per unit area) and the second term denotes the power radiated from the sample per unit area ($W$ is the sample thickness and $E_g$ is the band gap).\\

The effective thermal conductance depends on several factors like the conductivity and thickness of the materials involved, interface thermal resistance, etc \cite{incropera1996fundamentals}. For example, if the sample were placed between thin glass plates of thickness $W_G$ and in the absence of any interface thermal resistances, we have $G_T=2k_G/W_G$, where $k_G$ is the thermal conductivity of glass. It is evident that for large $G_T$, the sample will remain at $T=T_{amb}$. Further, air being a poor conductor, there could be significant interface thermal resistance between the sample and air. \\

Eq. \ref{eq:thermal}b re-states the heat flux balance condition in terms of the PLQE. Here, $E_{ph}=hc/\lambda$ denotes the energy of incident photons. Eq. \ref{eq:thermal}b  indicates that $T$ increases for a reduction in $PLQE$. Curiously, an increase in $T$ naturally leads to a decrease in $PLQE$ through the temperature dependence of parameters like $k_2$. Hence, as shown in Fig. \ref{schematic}, this is a positive feedback system which might lead to thermal runaways thus damaging the sample unless proper care is taken to ensure thermal dissipation pathways are enough and sufficient.\\

We note that eq. \ref{eq:thermal} assumes that the entire sample is at the same temperature. However, this assumption can be easily relaxed, if necessary, through compact models and an analogy with electrical circuits or through the numerical solution of  Fourier heat equation itself. This will not be attempted in this manuscript as our main aim is to unravel the essential mechanisms behind the reduction in PLQE at high excitation levels. In addition, eq. \ref{eq:thermal} also assumes that the entire incident photons are absorbed by the sample. Any reflection loss or photon re-absorption can be easily incorporated in the same formalism.\\ 

\section*{Results}
\begin {figure} [h]
  \centering
    \includegraphics[width=0.45\textwidth]{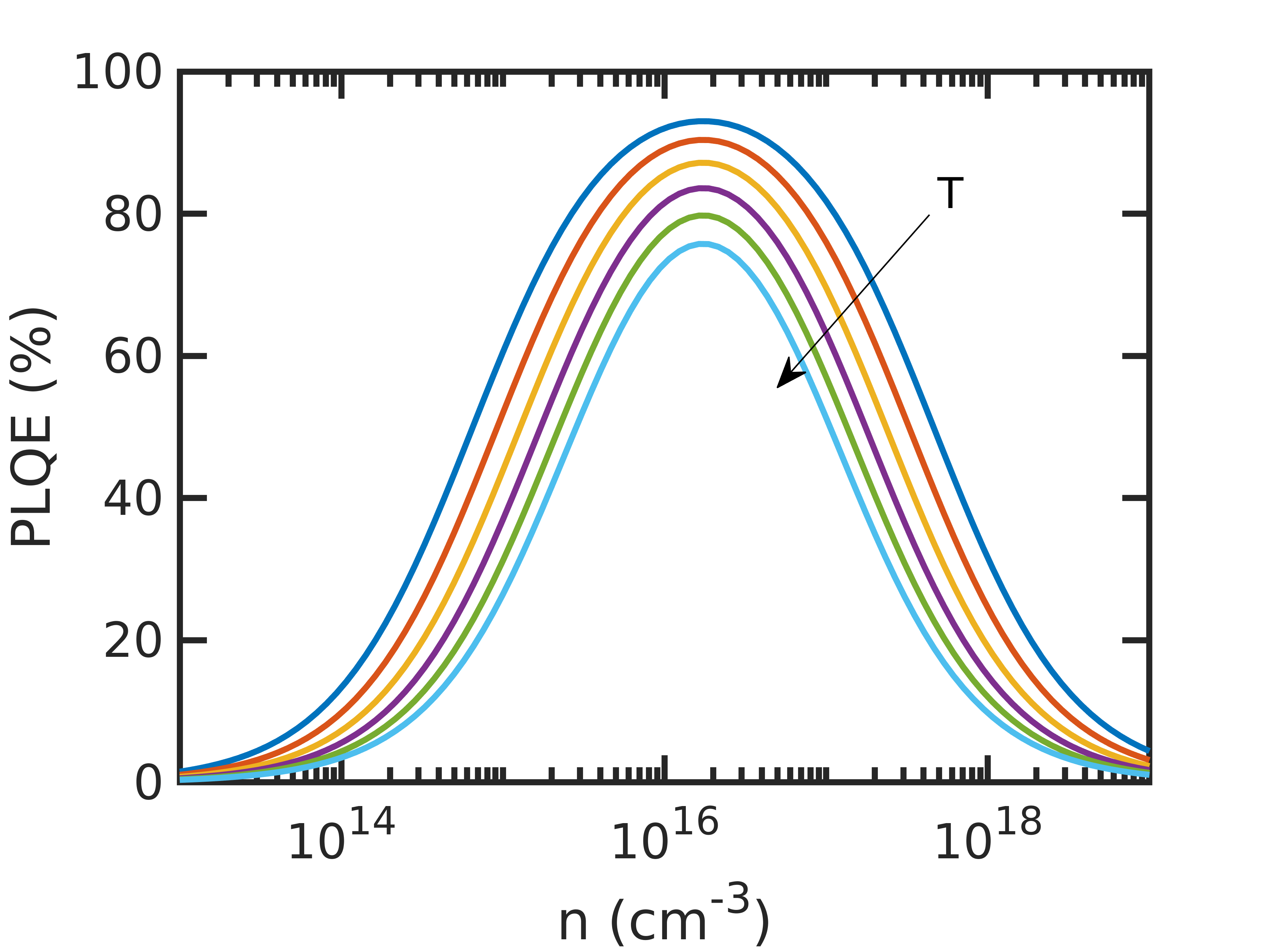}
    \caption{\textit{Variation of PLQE as a function of $n$ and $T$. Here, $k_2$ is assumed to be temperature dependent while $k_1$ and $k_3$ are assumed to be invariant with temperature. 
     }}
\label{PLQE_T}
\end{figure}

The temperature dependence of the $k_{2}$ is shown in Figure \ref{k2_T}. The key insight is that $k_{2}$ scales exponentially with an effective activation energy over the temperature range of interest ($300-350\, \mathrm{K}$).  Before we discuss the impact of self heating and the predictions of our proposed model, it is instructive to assess the parametric impact of temperature variation on the PLQE, as shown in Fig. \ref{PLQE_T}. Here, we assume that both $k_1$ and $k_3$ are independent of temperature while the $k_2$ varies as eq. \ref{eq:k2_T} with $E_A=150 \, \mathrm{meV} $ (and $k_{2}=5\times 10^{-11} \, \mathrm{cm^3/s}$ at $T=300\, \mathrm{K}$). \\

For a given set of parameters (i.e., at a given $T$), the PLQE vs. log(n) is symmetric about its maximum value. The luminescence droop or the roll off in PLQE beyond its maximum value (see Fg. \ref{PLQE_T}) is due to the increased contribution by the Auger recombination. However, in PLQE measurements, the change in $n$ is due to a change in illumination intensity $I$. As $I$ increases, the temperature $T$ of the sample increases due to thermal dissipation. Accordingly, the locus of PLQE vs. $n$ would shift along the family of curves shown in Fig. \ref{PLQE_T}. This results in a marked skew in its characteristics, especially for large $n$ - rather, the efficiency droop is expected to be more pronounced due to the temperature dependent reduction in $k_2$ in addition to the increase in Auger recombination (i.e., as $n$ increases).  \\

\begin {figure} [h]
  \centering
    \includegraphics[width=0.45\textwidth]{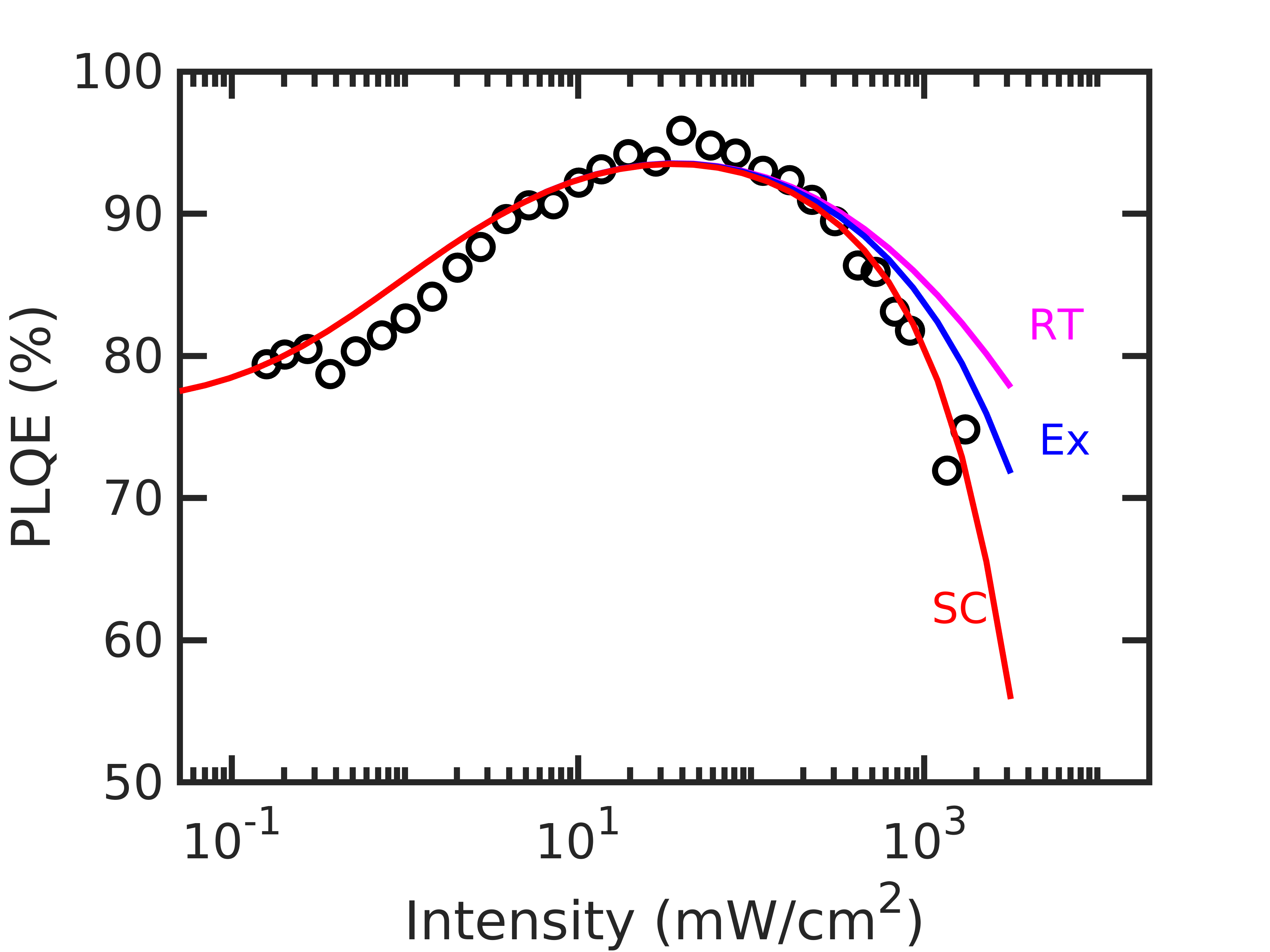}
    \caption{\textit{Comparison of model predictions (solid lines) with experiments (symbols). The experimental data is from ref. \cite{Li2024}. The data set marked as 'RT' denotes model predictions with $T=T_{amb}$. The results marked as 'Ex' denotes model predictions with temperature dependence only for excitonic contribution while the data set marked as 'SC' denotes the model predictions with temperature dependence for both $k_{2,rad}$ and excitons. All models yield same PLQE estimates for low $I$, while self-heating induced efficiency droop is  clearly evident for large $I$.
     }}
\label{exp_T}
\end{figure}

Figure \ref{exp_T} shows the comparison of model predictions with experimental results. The set of equations \ref{eq:sim_rate}-\ref{eq:thermal} is numerically solved (Newton method) under various conditions. The symbols in Fig. \ref{exp_T} are experimental results from literature which reported a record PLQE of 96\% for 3D perovskites and the solid lines represent numerical simulations (this work). The curve denoted as 'RT' indicates model predictions assuming $T=T_{amb}=300\, \mathrm{K} $ (rather, with very large $G_T$). While the comparison with experimental trends is good for small $I$, the model predictions do not accurately anticipate the efficiency droop for large $I$. The curve indicated as 'Ex' shows the impact due to  the temperature dependence of excitonic component. Here, temperature dependence for excitons is explicitly taken into account while the $k_{2,rad}$ is assumed to be temperature independent. Evidently, as the $T$ increases with $I$, the excitonic contribution to Photoluminescence decreases due to an increase in $n_{ep}$. This leads to a reduction in $k_2$ and hence an increase in the Auger recombination thus leading to efficiency droop.\\

The data set marked as 'SC' in Fig. \ref{exp_T} indicates self-consistent PLQE estimates with both $k_{2,rad}$ and excitonic contributions being temperature dependent (parameters are mentioned in Suppl. Mat.). A quick comparison between the data sets 'RT' and 'SC' indicates that the efficiency droop is more due to the self heating and temperature dependent reduction in $k_2$ which leads to increased contribution by Auger recombination. Our earlier analysis indicated that a reduction in PLQE to 50\% of its maximum value requires about an order of magnitude change in $n$ and hence about 2 orders of magnitude change in $I$. Aided by the self heating induced reduction in $k_2$ (say, by a factor of 2), the same change in PLQE (i.e., from 90\% to 50\% of the peak value) is possible due to an order of magnitude change in $I$. This clearly indicates the influence of positive feedback in the luminescence efficiency droop in perovskites. \\

\section*{Discussions}
\begin {figure} [h]
  \centering
    \includegraphics[width=0.45\textwidth]{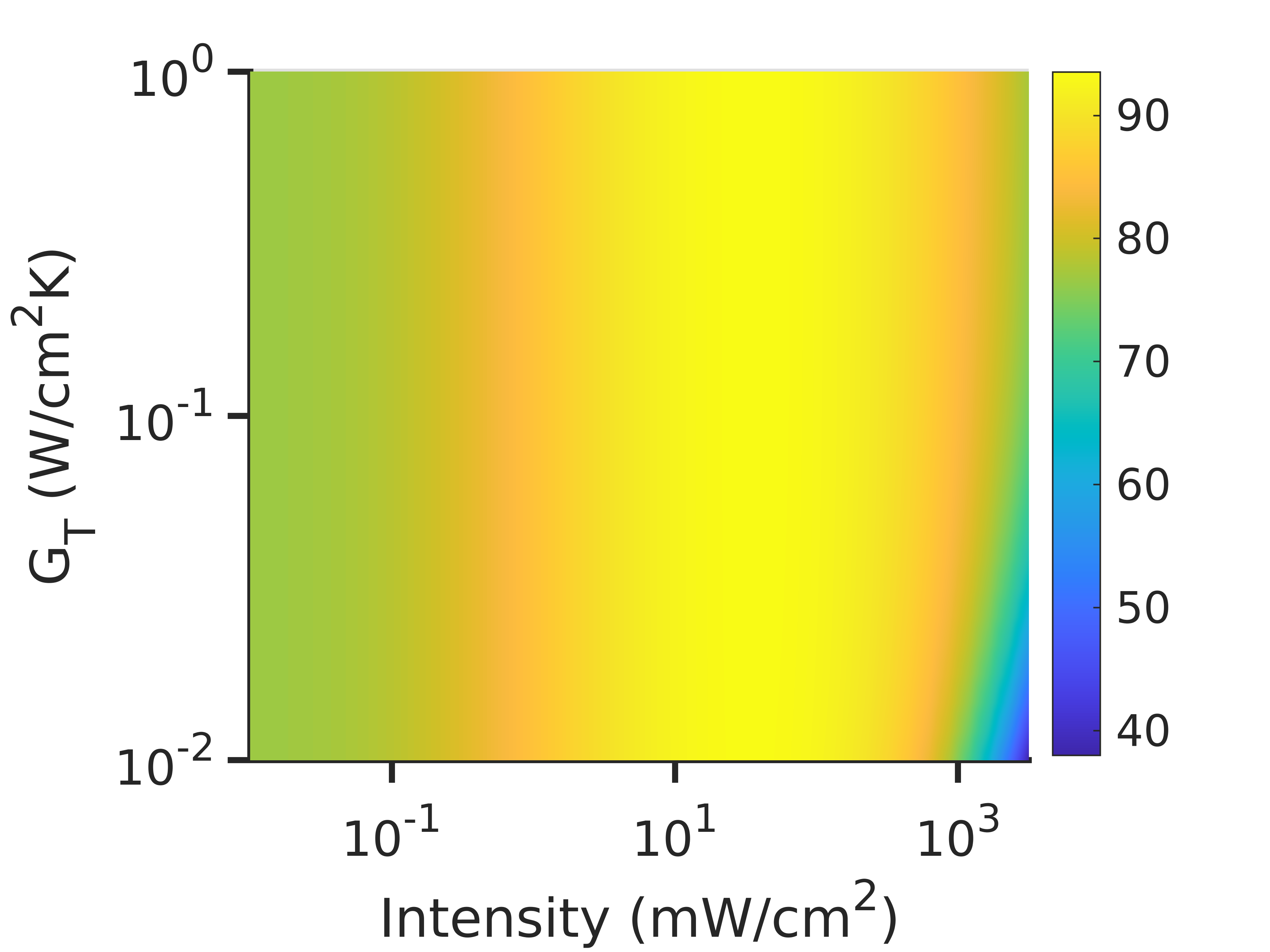}
    \caption{\textit{Influence of the effective thermal conductance and intensity on the PLQE (shown as color plot) of Perovskites. The parameters used are listed in Suppl. Mat.
     }}
\label{PLQE_GI}
\end{figure}

Equation \ref{eq:thermal} identifies the key factors that influence the efficiency droop in perovskites. An increase in $T$ coupled with the reduction in $k_2$ is the dominant factor. Accordingly, the key associated variables are $G_T$, $E_A$, and $E_{ph}$. Figure \ref{PLQE_GI} shows the influence of thermal conductance ($G_T$)  and illumination intensity on the temperature and hence PLQE of the sample. For low $G_T$ and large $I$, the sample temperature increases significantly thus lowering the PLQE. On the other hand, with large $G_T$, the model predictions indicate that the sample will remain close to ambient conditions and hence a much lower efficiency droop is expected. Rather, the device packaging should be such that the $G_T>0.1 \, \mathrm{W/cm^2K}$ to ensure that the efficiency loss due to thermal issues is minimized.\\

The parameter $k_2$ has a non-trivial influence on PLQE. It is obvious from eq. \ref{eq:sim_PLQE} that a large $k_2$ is needed to achieve good PLQE. However, Fig \ref{k2_T} and eqs. \ref{eq:ex}-\ref{eq:k2_T} indicates that excitonic effects, band tail states, etc. contribute to the radiative recombination rate. In fact, the detailed balance limit $k_{2,SQ}$  is much lower in magnitude compared to $k_{2,RS}$. While larger $k_2$ is beneficial, we find that $E_A$ also increases with $k_2$ (see Fig. \ref{k2_T}). An increase in $E_A$  results in rapid decrease of $k_2$ with temperature, which aids to increase the temperature of the sample through the positive feedback. Fig. S1 of Suppl. Mat. shows the PLQE droop as a function of $E_A$ and $G_T$ (i.e., the reduction in PLQE as $I$ is increased from $100 \, \mathrm{mW/cm^2}$ to $1000 \, \mathrm{mW/cm^2}$). As predicted by eq. \ref{eq:thermal}, the droop is minimized for small $E_A$ and large $G_T$. For such conditions, the self-heating is also minimized (see Fig. S2 of Suppl. Mat.).  These results indicate the need to accurately characterize the temperature dependence of $k_2$. An ideal material for LED applications should have a large $k_2$ with very small $E_A$.\\ 

Our analysis identifies the optimal measurement conditions to achieve the best PLQE for a sample. Note that the PLQE characteristics could be non-significantly influenced by the wavelength of the incident laser illumination (see eq. \ref{eq:thermal}b). As $\lambda$ reduces, $E_{ph}$ increases and hence the thermalization loss per incident photon increases. This leads to an increase in $T$ which lowers the PLQE for a given $I$. Hence, it is evident that the energy of the incident photons should be almost the same as the band gap of the semiconductor in PLQE measurements. This reduces thermalization loss and hence the EQE roll-off at high intensities.\\

The insights shared in this manuscript are of significant importance towards back extraction of key parameters like $k_3$. If the self-heating issues are neglected, it requires a much larger contribution from Auger recombination to account for the rapid decrease in PLQE beyond its peak value (see Fig. \ref{exp_T}). Accordingly, strategies to back extract $k_3$ using the efficiency droop characteristics need to account for self-heating aspects. The model developed also allows exploration of the temperature dependent degradation/phase segregation in perovskite LEDs and solar cells. During long term operation, unless the thermal design is appropriate, the temperature of the sample/device could increase which leads to material degradation and/or phase segregation. This could lead to an increase in $k_1$ due to an increase in recombination centers and hence a reduction in PLQE. Phase segregation could lead to spatial heterogenity \cite{slotcavage2016light,braly2017current,abhimanyu_JAP2021} where carrier accumulation in low band gap regions could lead to increased recombination and hence reduction in PLQE. Such long term degradation could be addressed through the formalism developed in this manuscript. \\

\section*{Conclusions}
In summary, here we report a self-consistent analysis of the luminescence efficiency droop in perovskites. Our analysis indicates that self-heating is significant under large illumination intensities. Under such conditions, through a positive feedback mechanism, the radiative recombination rate decreases which leads to an increase in Auger recombination thus leading to a droop in the luminescence efficiency. These insights are of broad interest towards the thermal design and packaging of  perovskite based light emitting diodes and solar cells. 

\section*{Acknowledgement}
Authors acknowledge National Center for Photovoltaics Research and Education (NCPRE), Indian Institute of Technology Bombay.\\

\section*{Supplementary Material}
Dependence of efficiency droop and sample temperature on the $E_A$, and $G_T$ is provided in Suppl. Mat. The same document also lists the set of parameters used in this manuscript.\\

\section*{References}
\bibliography{apssamp}

\end{document}